\pdfoutput=1
\documentclass[11pt]{article}
\usepackage[T1]{fontenc}
\usepackage[margin=1in]{geometry}
\usepackage{amsmath,amssymb,amsthm,booktabs,array}
\usepackage[shortlabels]{enumitem}
\usepackage{pgfplots}
\usepgfplotslibrary{groupplots}
\pgfplotsset{compat=1.18}
\usepackage[hidelinks]{hyperref}

\setlist{itemsep=1pt,topsep=3pt}

\newtheorem{theorem}{Theorem}[section]
\newtheorem{lemma}[theorem]{Lemma}
\newtheorem{proposition}[theorem]{Proposition}
\theoremstyle{remark}
\newtheorem{remark}[theorem]{Remark}

\newcommand{\poly}{\mathrm{poly}}
\newcommand{\COL}{\textup{\textsc{Col}}}

\title{Polynomial-Space Quantum Algorithms from Classical Branching: Colouring, Domination and Monotone Local Search}
\hypersetup{pdftitle={Polynomial-Space Quantum Algorithms from Classical Branching: Colouring, Domination and Monotone Local Search},pdfauthor={Sebastian Senge}}
\author{Sebastian Senge\\ Independent Researcher}
\date{}

\begin{document}
\maketitle

\begin{abstract}
We give quantum algorithms for exponential-time problems on $n$ vertices or elements with bounded error, polynomial space and no quantum random access memory. Our results are these:
\begin{itemize}[leftmargin=1.5em]
\item 4-Colouring can be solved in quantum time $O(1.3032^n)$, from the polynomial-space branching algorithm of~\cite{WGJ24} with terminal algorithms of~\cite{YS26}. With~\cite{WGJ24} alone the bound is $O(1.3100^n)$.
\item For $k \in \{5, 7, 8, 9, 10, 11\}$ and $13 \le k \le 18$, $k$-Colouring can be solved in quantum time below the previous best bound in this model to our knowledge, from $O(1.4518^n)$ for $k = 5$ to $O(1.9233^n)$ for $k = 18$.
\item Dominating Set can be solved in quantum time $O(1.2235^n)$, Independent Dominating Set in $O(1.1583^n)$, and Edge Dominating Set and Minimum Maximal Matching in $O(1.1501^n)$.
\item A quantum form of monotone local search: if the extension problem has a backtracking tree with $c^k$ nodes up to polynomial factors, the problem is solved in quantum time $(2 - 1/c)^{n/2}$ up to polynomial factors. This gives Feedback Vertex Set in $O(1.3081^n)$, Subset and Group Feedback Vertex Set in $O(1.3229^n)$, Cluster Vertex Deletion in $O(1.2033^n)$, 3-Hitting Set in $O(1.2405^n)$, Proper Interval Vertex Deletion in $O(1.3541^n)$, Multicut on Trees in $O(1.1647^n)$ and Min-Ones $d$-SAT in $O^*((2-1/d)^{n/2})$. The approximate form gives quantum exponential-time approximations below the quantum exact bounds, such as a $1.1$-approximation for Vertex Cover in $O(1.0555^n)$ and for Feedback Vertex Set in $O(1.2161^n)$.
\item The exponents for $k$-colouring with $k \ge 6$ come from an evaluation of the reduction recursion whose values bound the running time of an explicit algorithm, computed in floating point with a margin.
\end{itemize}
To our knowledge, the previous best bounds in this model are $O(1.3231^n)$ for $k = 4$ and $O(1.4695^n)$ to $O(1.9366^n)$ for the other values of $k$ above~\cite{SM22}, and $O(2^{n/2})$ for Dominating Set~\cite{WCCV23}. For the other domination problems we know of no stated quantum bound; the walk on the enumeration of maximal independent sets gives $O^*(3^{n/6}) \subset O(1.2010^n)$.
\end{abstract}

\section{Introduction}

Deciding whether a graph has a proper colouring with $k$ colours is one of the basic NP-complete problems. For $k = 3$ the algorithm of~\cite{BE05} is a branching algorithm in polynomial space with running time $O^*(1.3289^n)$. For $k = 4$ the algorithm of~\cite{WGJ24} is a branching algorithm in polynomial space with running time $O(1.7159^n)$. The recent work~\cite{YS26} gives $O^*(1.3204^n)$ and $O^*(1.6975^n)$, with dynamic programming over tree decompositions, which uses exponential space. For larger $k$, colouring reduces to smaller $k$ through maximal independent sets~\cite{Law76,Bys04} and through balanced splits of the colour classes.

Quantum search applies to branching algorithms, as noted in~\cite{Fur08}, and gives the square root of the classical time of~\cite{BE05}, so 3-colouring in quantum time $O(1.1528^n)$. The work~\cite{SM22} combines this with the reductions above and gives polynomial-space quantum algorithms without quantum random access memory for every $k \le 20$ below $O(2^n)$. Their Theorem~9 takes the exponent of 3-colouring as input. With quantum random access memory, the chromatic number can be computed in time $O(1.7956^n)$~\cite{GL24}, which is outside the model of this paper.

For domination problems, the best polynomial-space classical algorithms are branch and reduce algorithms analysed by measure and conquer~\cite{RB11,RB12,BEP10,BDEP13,Iwa11,XN14}. For Dominating Set, a quantum bound of $O(2^{n/2})$ queries was stated in~\cite{WCCV23}. For many vertex deletion and hitting set problems the best classical bounds come from monotone local search~\cite{FGLS19}, which samples part of a solution at random and completes it with a parameterized algorithm.

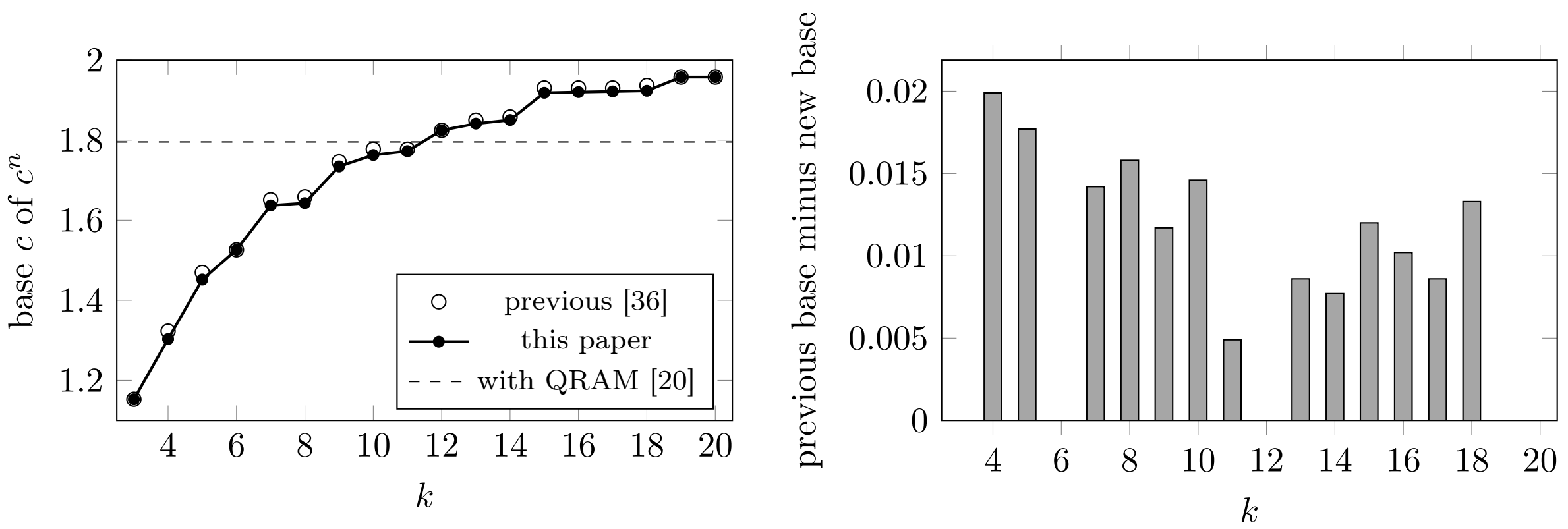
\begin{figure}[t]
\centering
\begin{tikzpicture}
\begin{groupplot}[group style={group size=2 by 1, horizontal sep=2.1cm}, width=0.47\textwidth, height=5.2cm,
  xmin=2.5, xmax=20.5, xtick={4,6,...,20}, xlabel={$k$}, tick label style={font=\small}, label style={font=\small}]
\nextgroupplot[ylabel={base $c$ of $c^n$}, ymin=1.1, ymax=2.0, legend pos=south east, legend style={font=\scriptsize}]
\addplot[only marks, mark=o, mark size=2pt] coordinates {(3,1.1528) (4,1.3231) (5,1.4695) (6,1.5261) (7,1.6511) (8,1.6585) (9,1.746) (10,1.7775) (11,1.7775) (12,1.8246) (13,1.8499) (14,1.858) (15,1.9303) (16,1.9303) (17,1.9303) (18,1.9366) (19,1.9575) (20,1.9575)};
\addlegendentry{previous~\cite{SM22}}
\addplot[thick, mark=*, mark size=1.3pt] coordinates {(3,1.1528) (4,1.3032) (5,1.4518) (6,1.5261) (7,1.6369) (8,1.6427) (9,1.7343) (10,1.7629) (11,1.7726) (12,1.8246) (13,1.8413) (14,1.8503) (15,1.9183) (16,1.9201) (17,1.9217) (18,1.9233) (19,1.9575) (20,1.9575)};
\addlegendentry{this paper}
\addplot[dashed, domain=2.5:20.5] {1.7956};
\addlegendentry{with QRAM~\cite{GL24}}
\nextgroupplot[ylabel={previous base minus new base}, ymin=0, ybar, bar width=5pt, ytick={0,0.005,0.01,0.015,0.02}, scaled y ticks=false,
  yticklabel style={/pgf/number format/fixed, /pgf/number format/precision=3}]
\addplot[fill=black!35] coordinates {(3,0.0) (4,0.0199) (5,0.0177) (6,0.0) (7,0.0142) (8,0.0158) (9,0.0117) (10,0.0146) (11,0.0049) (12,0.0) (13,0.0086) (14,0.0077) (15,0.012) (16,0.0102) (17,0.0086) (18,0.0133) (19,0.0) (20,0.0)};
\end{groupplot}
\end{tikzpicture}
\caption{Quantum time $c^n$ for $k$-colouring with bounded error, polynomial space and no quantum random access memory. Left: the bases of~\cite{SM22} and of Table~\ref{tab:results}, and the bound $1.7956^n$ of~\cite{GL24}, which uses quantum random access memory. Right: the decrease of the base for each $k$; it is zero for $k = 3, 6, 12, 19, 20$.}
\label{fig:bases}
\end{figure}

\paragraph{Our contribution.} We show that the 4-colouring algorithm of~\cite{WGJ24} is a backtracking tree in the sense of~\cite{Mon18}, which gives 4-colouring in quantum time $O^*(1.7159^{n/2}) \subset O(1.3100^n)$. The terminal algorithms of~\cite{YS26} that use no dynamic programming over decompositions keep this form, and with 3-colouring from~\cite{BE05} they give at most $p(n)\,1.6981^n$ nodes, so quantum time $O^*(1.6981^{n/2}) \subset O(1.3032^n)$. We then evaluate the recursion of~\cite{SM22} with 4-colouring as a second base case. The evaluation bounds the recursion from above on pieces of a partition, with all size limits rounded up to a grid, and we show that the algorithm can be run with these rounded limits. So the values are upper bounds on the time of an explicit algorithm. Table~\ref{tab:results} and Figure~\ref{fig:bases} show the results. In Section~\ref{sec:dom} we show that the polynomial-space algorithms of~\cite{RB11,RB12,BEP10} for domination problems are backtracking trees in their decision form, which gives the bounds of Table~\ref{tab:dom}. In Section~\ref{sec:mls} we give a quantum form of monotone local search, with amplitude amplification over the samples and the walk of~\cite{Mon18} on the extension algorithm, and apply it to the problems of Table~\ref{tab:mls}.

\begin{table}[h]
\centering
\small
\renewcommand{\arraystretch}{1.15}
\begin{tabular}{@{}rlll@{}}
\toprule
$k$ & Previous best bound~\cite{SM22} & With~\cite{WGJ24} alone & This paper, with~\cite{WGJ24,YS26}\\
\midrule
4 & $1.3231^n$ & $1.3100^n$ & $\mathbf{1.3032^n}$\\
5 & $1.4695^n$ & $1.4578^n$ & $\mathbf{1.4518^n}$\\
7 & $1.6511^n$ & $1.6418^n$ & $\mathbf{1.6369^n}$\\
8 & $1.6585^n$ & $1.6481^n$ & $\mathbf{1.6427^n}$\\
9 & $1.7460^n$ & $1.7383^n$ & $\mathbf{1.7343^n}$\\
10 & $1.7775^n$ & $1.7679^n$ & $\mathbf{1.7629^n}$\\
11 & $1.7775^n$ & $1.7726^n$ & $\mathbf{1.7726^n}$\\
13 & $1.8499^n$ & $1.8442^n$ & $\mathbf{1.8413^n}$\\
14 & $1.8580^n$ & $1.8531^n$ & $\mathbf{1.8503^n}$\\
15 & $1.9303^n$ & $1.9225^n$ & $\mathbf{1.9183^n}$\\
16 & $1.9303^n$ & $1.9226^n$ & $\mathbf{1.9201^n}$\\
17 & $1.9303^n$ & $1.9242^n$ & $\mathbf{1.9217^n}$\\
18 & $1.9366^n$ & $1.9279^n$ & $\mathbf{1.9233^n}$\\
\bottomrule
\end{tabular}
\caption{Quantum time for $k$-colouring with bounded error, polynomial space and no quantum random access memory. All values are rounded up. The previous bounds are the best known to us. For $k = 3, 6, 12, 19, 20$ the bounds of~\cite{SM22} stand.}
\label{tab:results}
\end{table}

\paragraph{Scope.} The last column of Table~\ref{tab:results} uses terminal algorithms of~\cite{YS26}, an arXiv paper that is, to our knowledge, unrefereed. The column with~\cite{WGJ24} alone does not depend on it. The bound $O^*(1.3217^n)$ for 3-colouring in the arXiv paper~\cite{Mei23} is not used. Its linear program, as printed, has a larger optimum~\cite[Appendix~F]{YS26}, and Remark~\ref{rem:mei} states what follows from that program. For the rows $k = 4, 5$, the arithmetic of the closed forms is checked in Lean. The rows $k = 6$ to $18$ come from a floating point computation with a margin far above its rounding error, not from interval arithmetic. The bases of the domination problems come from numerically computed analyses in their sources, used as stated there. Numbers of theorems, lemmas and appendices of~\cite{SM22,YS26} refer to their arXiv versions. Code and Lean files are at \url{https://github.com/ssenge/LAIF}, in the folders \texttt{quantum-graph-colouring}, \texttt{quantum-dominating-set}, \texttt{quantum-monotone-local-search}, \texttt{quantum-approximate-monotone-local-search} and \texttt{quantum-vertex-cover}, built with Lean v4.36.0-rc1 and Mathlib at revision {\sloppy\texttt{d875a8de5222f62cf1c2fafc7984588adadc990a}.\par}

\section{Preliminaries}

For $\delta \in [0,1]$ let $h(\delta) = -\delta\log\delta - (1-\delta)\log(1-\delta)$, with logarithms to base 2. $O^*$ hides factors polynomial in $n$.

\paragraph{Backtracking trees.} A backtracking tree is given by a root instance and a procedure that, in time $\poly(n)$, decides for an instance whether it is a leaf, and if so whether it is marked, or else lists its children. The depth is at most $\poly(n)$. Given a number $T$ computable in time $\poly(n)$, the quantum walk of~\cite{Mon18} decides whether a tree with at most $T$ nodes has a marked leaf, with bounded error, in time $O(\sqrt{T}\,\poly(n))$ and polynomial space. Its circuit depends only on $n$ and $T$, so it can be used inside quantum search.

\paragraph{Maximal independent sets.} By~\cite{Bys04}, an $n$-vertex graph has at most $I(n,t) = O^*(2^{E(t/n)n})$ maximal independent sets of size $t$, for a concave piecewise linear function $E$ with $E(1/s) = (\log s)/s$, and they can be listed by a branching algorithm whose number of leaves is bounded by $I(n,t)$ in the form required by Lemma~4 of~\cite{SM22}.

\section{3-Colouring and 4-Colouring}

\paragraph{Constraint satisfaction.} In a $(3,2)$-constraint satisfaction instance, every variable has at most three values and every constraint forbids one pair of values on two variables. A variable with at most two values can be eliminated in polynomial time (Lemma~2 of~\cite{BE05}). The algorithm of Theorem~1 of~\cite{BE05} works by reductions with one call and by branching into a constant number of smaller instances, with polynomial work per call, and it solves its leaves by a bipartite matching. Its analysis bounds the number of recursive calls by $p(r)\,\Lambda^r$ on an instance with $r$ three-valued variables, where $\Lambda < 1.36444$ is the largest root of $x^5 = 2x + 2$. So it is a backtracking tree with at most this number of nodes.

\begin{lemma}\label{lem:three}
The 3-colouring algorithm of~\cite{BE05} is a backtracking tree with at most $p(n)\,\beta^n$ nodes on graphs with $n$ vertices, where $\beta = (2^3\, 3^4\, \Lambda^{24})^{1/49} < 1.3289$ and $p$ is a fixed polynomial.
\end{lemma}

\begin{proof}
The algorithm first applies reductions and the branching on cycles of vertices of degree three, each an OR over smaller instances. It then chooses a forest in polynomial time, and tries colourings of a set of vertices, one vertex at a time, with a node that stores the forest, the order and the partial colouring. Each such colouring leaves a $(3,2)$-constraint satisfaction instance. Theorem~2 of~\cite{BE05} bounds the product of these counts with the size of the search trees of the last step, which gives the bound on the nodes, up to the polynomial depth. This is the form used in Lemma~6 of~\cite{SM22}.
\end{proof}

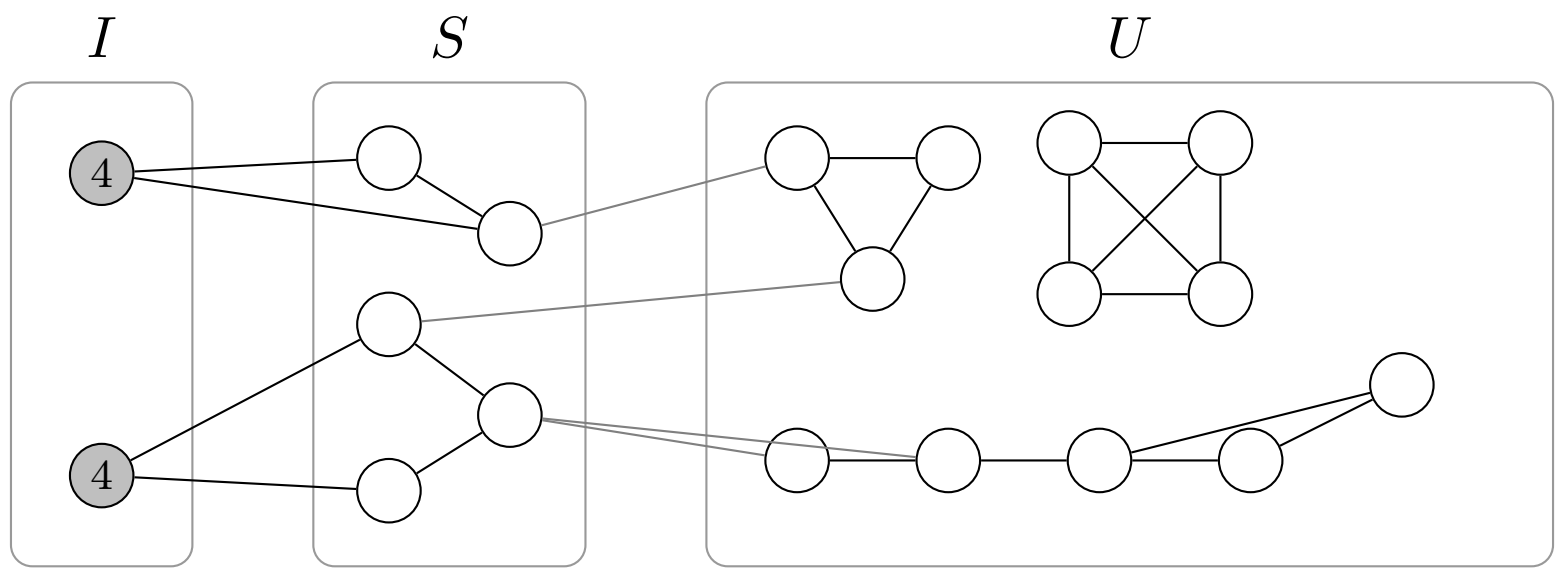
\begin{figure}[t]
\centering
\begin{tikzpicture}[v/.style={circle, draw, inner sep=0pt, minimum size=4.2mm, font=\scriptsize}, every edge/.style={draw}]
\draw[rounded corners, black!40] (-0.6,-0.6) rectangle (0.6,2.6); \node at (0,2.9) {$I$};
\draw[rounded corners, black!40] (1.4,-0.6) rectangle (3.2,2.6); \node at (2.3,2.9) {$S$};
\draw[rounded corners, black!40] (4.0,-0.6) rectangle (9.6,2.6); \node at (6.8,2.9) {$U$};
\node[v, fill=black!25] (i1) at (0,2) {4}; \node[v, fill=black!25] (i2) at (0,0) {4};
\node[v] (s1) at (1.9,2.1) {}; \node[v] (s2) at (2.7,1.6) {}; \node[v] (s3) at (1.9,1.0) {};
\node[v] (s4) at (2.7,0.4) {}; \node[v] (s5) at (1.9,-0.1) {};
\node[v] (a1) at (4.6,2.1) {}; \node[v] (a2) at (5.6,2.1) {}; \node[v] (a3) at (5.1,1.3) {};
\node[v] (b1) at (6.4,2.2) {}; \node[v] (b2) at (7.4,2.2) {}; \node[v] (b3) at (6.4,1.2) {}; \node[v] (b4) at (7.4,1.2) {};
\node[v] (c1) at (4.6,0.1) {}; \node[v] (c2) at (5.6,0.1) {}; \node[v] (c3) at (6.6,0.1) {}; \node[v] (c4) at (7.6,0.1) {}; \node[v] (c5) at (8.6,0.6) {};
\draw (i1) edge (s1) (i1) edge (s2) (i2) edge (s3) (i2) edge (s5) (s1) edge (s2) (s3) edge (s4) (s4) edge (s5);
\draw (a1) edge (a2) (a2) edge (a3) (a3) edge (a1);
\draw (b1) edge (b2) (b1) edge (b3) (b1) edge (b4) (b2) edge (b3) (b2) edge (b4) (b3) edge (b4);
\draw (c1) edge (c2) (c2) edge (c3) (c3) edge (c4) (c4) edge (c5) (c3) edge (c5);
\draw[black!50] (s2) edge (a1) (s3) edge (a3) (s4) edge (c1) (s4) edge (c2);
\end{tikzpicture}
\caption{An instance $P = (G, I, S)$ of 4-colouring. The vertices of $I$ have colour $4$, the vertices of $S \supseteq N(I)$ have colours in $\{1,2,3\}$, and the vertices of $U$ are free. The components of $G[U]$ at the top are a $K_3$ and a $K_4$. The component at the bottom is neither, so the instance is not terminal, and a branching rule of~\cite{WGJ24} acts on it.}
\label{fig:instance}
\end{figure}

\paragraph{The instances of 4-colouring.} We use the framework of~\cite{WGJ24}, with the notation of~\cite{YS26}. An instance is $P = (G, I, S)$, where $I$ is an independent set, $S \supseteq N(I)$ is disjoint from $I$, the vertices of $I$ get colour $4$, the vertices of $S$ get colours in $\{1,2,3\}$, and the vertices of $U = V \setminus (I \cup S)$ are free (Figure~\ref{fig:instance}). The instance is terminal if every component of $G[U]$ is a $K_3$ or a $K_4$. Let $n_3$ and $n_4$ count these components. The measure is
\[
\mu(P) = \omega_c |S| + \omega_{3}' n_3 + \omega_{4}' n_4 + \sum_{C} \sum_{v \in C} \omega_{\deg_C(v)},
\]
summing over the other components $C$ of $G[U]$, with $\omega_0 = 0$ and $\omega_i = 1$ for $i \ge 3$. If the weights satisfy Equation~(3) of~\cite{WGJ24}, then $\mu(G, \emptyset, \emptyset) \le n$.

A nonterminal instance is replaced by the branching rules of Section~3.2 of~\cite{WGJ24}. Each rule gives instances $P_j$ such that $P$ has a colouring if and only if some $P_j$ has one, and $|U|$ decreases in each $P_j$. Their number is a constant, except on a cycle component, where the rule takes one $P_j$ for each maximal independent set of the cycle. If the weights and $r$ satisfy Equations~(3) and~(6) to~(21) of~\cite{WGJ24}, then $\sum_j r^{\mu(P_j)} \le r^{\mu(P)}$, which is Lemma~5.7 of~\cite{YS26}. Every clique of a terminal instance may be assumed to contain a vertex of colour $4$, since one of its vertices can be recoloured with colour $4$ otherwise: it has no neighbour in $I$, and its neighbours in $S$ do not use this colour.

\paragraph{Terminal instances.} From a terminal instance we compute in polynomial time two maximal families of disjoint copies of $K_{1,3}$, one in $G[S]$ with vertex set $S_1$, and one with centres in $U$ and leaves in $S \setminus S_1$. Let $s_1 = |S_1|$, let $s'$ be the number of vertices of $S$ in neither family, and let $a_i$ and $b_i$ count the components $K_3$ and $K_4$ with exactly $i$ centres. Then $G[S \setminus S_1]$ has maximum degree at most two (Lemma~5.1 of~\cite{YS26}). We use these terminal algorithms, all of them OR trees whose steps take polynomial time and space.
\begin{enumerate}[(T1)]
\item Choose the vertex of colour $4$ in each clique and 3-colour the rest by Lemma~\ref{lem:three} (Lemma~5.2 of~\cite{YS26}). The size of the tree is $3^{n_3} 4^{n_4} \beta^{|S| + 2n_3 + 3n_4}$ up to polynomial factors.
\item Choose the colours of the centres of both families, and restrict the cliques without centres to lists of two colours, then solve the $(3,2)$-constraint satisfaction instance (Lemma~5.4 of~\cite{YS26}). The size is $3^{s_1/4} \Lambda^{s'} \prod_i g_{3,i}^{a_i} \prod_i g_{4,i}^{b_i}\, e^{o(n)}$, with $g_{3,0} = 9/2$, $g_{4,0} = 6$, $g_{m,1} = \Lambda^{m-1}(3 + \Lambda^3)$ and $g_{m,i} = \frac{6}{(4-i)!}(4 - i + i\Lambda^3)$ for $i \ge 2$. The factor $9/2$ comes from a family of list assignments on blocks of $O(\log n)$ triangles, chosen greedily in polynomial time (Lemma~5.3 of~\cite{YS26}).
\item Enumerate the colours of $S_1$ and of the cliques, then solve list colouring on the paths and cycles $G[S \setminus S_1]$ in polynomial time (Lemma~5.5 of~\cite{YS26}). The size is $24^{s_1/4} 18^{n_3} 24^{n_4}$ up to polynomial factors.
\item Enumerate the sets of at most $n_3/3$ triangles that miss one fixed colour of $\{1,2,3\}$, which suffices by the symmetry of these colours, restrict the domains as in Section~3.1 of~\cite{WGJ24}, and solve the $(3,2)$-constraint satisfaction instances. The size is $1.3645^{|S|} 4.837^{n_3} 6^{n_4}$ up to polynomial factors.
\end{enumerate}

\begin{theorem}\label{thm:four}
4-Colouring has a quantum algorithm with bounded error, polynomial space and no quantum random access memory, running in time $O^*(1.6981^{n/2}) \subset O(1.3032^n)$. With the terminal algorithm (T4) alone, the time is $O^*(1.7159^{n/2}) \subset O(1.3100^n)$.
\end{theorem}

\begin{proof}
The tree starts at $(G, \emptyset, \emptyset)$, branches by the rules above and, at a terminal instance, computes the numbers $s_1, s', a_i, b_i$ and runs the terminal algorithm with the smallest stated size. A node stores the instance and the choices made so far, and a choice among polynomially many values is one step. On a cycle component the maximal independent set is chosen one vertex at a time, keeping only choices that can still be completed, so each such node lies above some $P_j$ and these nodes add a polynomial factor. The leaves are the leaves of the $(3,2)$ trees, of the 3-colouring trees, or of the list colouring step, and a leaf is marked if it gives a colouring. So there is a marked leaf if and only if $G$ is 4-colourable, and the depth is polynomial.

For the size, write $x = (s_1, s', a_0, \dots, a_3, b_0, \dots, b_4)$, and write each terminal size as $\exp(R_i x)$ with a row $R_i$ read off from (T1), (T2) and (T3). Lemma~5.8 of~\cite{YS26} gives the following. If $\eta \ge 0$ with $\sum_i \eta_i = 1$ and $\ln r\,(\omega_c |S| + \omega_3' n_3 + \omega_4' n_4) \ge \sum_i \eta_i R_i x$ for all $x \ge 0$, then $\min_i R_i x \le \mu(P)\ln r$ at every terminal instance. Since $\sum_L r^{\mu(L)} \le r^{\mu(G, \emptyset, \emptyset)} \le r^n$ over the terminal instances $L$, the number of nodes is at most $r^n$ times a polynomial and a factor $e^{o(n)}$. All constraints hold at $r = 1.69804$ with
\[
\omega_1 = 0.9342533916,\ \ \omega_2 = 1,\ \ \omega_c = 0.5593776725,\ \ \omega_3' = 2.9704494860,\ \ \omega_4' = 3.5579545637,
\]
and $\eta = (0.0822508347,\, 0.8778792406,\, 0.0398699247)$. Our script checks each constraint with 60 digits and a margin of $10^{-9}$. So there are constants $C$ and $a$ with at most $T_n = C\, n^a\, 1.6981^n$ nodes on every graph with $n$ vertices. The walk of~\cite{Mon18} with this bound decides 4-colourability in time $O^*(1.6981^{n/2})$, the same for every input of size $n$, and $1.6981 \le 1.3032^2$. For (T4) alone, the parameters of Table~2 of~\cite{WGJ24} satisfy its Equations~(3) to~(21), where~(4) and~(5) bound the size of (T4) by $r^{\mu(P)}$ at $r = 1.7159$. This is the proof of Theorem~8 of~\cite{WGJ24}, and it gives the bound $T_n$ with base $1.7159$, and $1.7159 \le 1.3100^2$.

A colouring follows by self-reduction. Add a $K_4$ on four new vertices. For each vertex $v$ in turn, merge $v$ with one of the new vertices, and keep the first merge that leaves a 4-colourable graph. The graph has at most $n + 4$ vertices throughout, so this costs a factor $O(n)$ and a constant factor in the time.
\end{proof}

The classical bound $O^*(1.6975^n)$ of~\cite{YS26} uses its 3-colouring algorithm in (T1), which needs exponential space, while its (T2) and (T3) do not. So the classical form of the tree above is a polynomial-space 4-colouring algorithm in time $O(1.6981^n)$, implied by~\cite{WGJ24,YS26}.

\begin{remark}\label{rem:mei}
The work~\cite{Mei23} gives a 3-colouring algorithm through a new choice of the forests and a linear program over their structure. The running time stated there is $O^*(1.3217^n)$, while the program as printed has optimum $1.32702$~\cite[Appendix~F]{YS26}. This does not bound the running time from below, and further constraints on the forests may lower the optimum. The algorithm has the steps of Lemma~\ref{lem:three}. Its vertices are coloured in the order of its forests, so it is a backtracking tree. If the constraints of that program hold for its forests, it has at most $p(n)\,1.32703^n$ nodes. In that case the same computation gives quantum time $O(1.1520^n)$ for 3-colouring, $O(1.5255^n)$ for $k = 6$, $O(1.8242^n)$ for $k = 12$ and $O(1.9574^n)$ for $k = 19$, and $O(1.7722^n)$ for $k = 11$, with the other rows of Table~\ref{tab:results} unchanged. If $1.3217$ holds, the 3-colouring bound becomes $O(1.1497^n)$.
\end{remark}

\section{\texorpdfstring{$k$}{k}-Colouring}

\subsection{The recursion}

Let $\COL(G, k, \mu)$ be a procedure that answers yes if the vertex set of $G$, of size $n$, splits into $k$ independent sets of size at most $\mu n$ each, answers no if $G$ is not $k$-colourable, and may answer either otherwise. Then $\COL(G, k, 1)$ decides $k$-colourability. Calling $\COL$ with a larger $\mu$ only weakens the first condition, so it is always correct to replace $\mu$ by a larger value.

Theorem~9 of~\cite{SM22} gives such a procedure for $k \ge 4$ from two reductions, each with quantum search over its choices.
\begin{itemize}
\item \textbf{Reduction 1.} Search over the maximal independent sets $I$ of size $t \ge n/k$, and call $\COL(G - I, k - 1, 1)$. In a colouring with $k$ classes the largest class can be extended to a maximal independent set, which has size at least $n/k$.
\item \textbf{Reduction 2.} Fix $2 \le k' \le k/2$. Search over the sets $S$ of size $t$ with $k'n/k \le t \le \min(n, \mu k' n)$, and answer yes if $\COL(G[S], k', \mu n/t)$ and $\COL(G - S, k - k', \lfloor t/k' \rfloor/(n - t))$ both answer yes. For $t = n$ the graph $G - S$ is empty and the second call is omitted. If $S$ is the union of the $k'$ largest classes of a colouring with classes of size at most $\mu n$, then $S$ has this size, its classes have size at most $\mu n$, and the remaining $k - k'$ classes are at most the average $t/k'$ of the $k'$ largest ones.
\end{itemize}
Let $f_3$ and $f_4$ be exponents of 3-colouring and 4-colouring, let $d_3 = f_3$ and $d_1 = d_2 = 0$, and for $k \ge 4$ and $\mu \in [1/k, 1]$ let
\begin{align}
d_k(\mu) = \min\Bigl(&\max_{1 \le s \le k} \frac{\log s}{2s} + \Bigl(1 - \frac1s\Bigr) d_{k-1}(1),\label{eq:r1}\\
&\min_{2 \le k' \le k/2}\; \max_{\delta \in [k'/k,\, \min(1, \mu k')]} \frac{h(\delta)}{2} + \max\Bigl(\delta\, d_{k'}\bigl(\min(1, \tfrac{\mu}{\delta})\bigr),\label{eq:r2}\\
&\hspace{16em} (1-\delta)\, d_{k-k'}\bigl(\min(1, \tfrac{\delta}{k'(1-\delta)})\bigr)\Bigr)\Bigr).\notag
\end{align}
Line~\eqref{eq:r1} is the time of Reduction~1: quantum search over $I(n,t)$ sets costs $2^{E(\delta)n/2}$ for $\delta = t/n$, the maximum of $E(\delta)/2 + (1-\delta)d_{k-1}(1)$ over $\delta \in [1/k, 1]$ is attained at some $\delta = 1/s$, since $E$ is piecewise linear with breakpoints at these points (Corollary~3 of~\cite{SM22}), and $E(1/s) = (\log s)/s$. The term $s = 1$ is $0$ and is included only to make the range simple; \cite{SM22} uses $2 \le s \le k$. Line~\eqref{eq:r2} is the time of Reduction~2, with $\binom{n}{t} = O^*(2^{h(\delta)n})$ sets. By Theorem~9 of~\cite{SM22}, $k$-colouring has a quantum algorithm in time $O^*(2^{d_k(1)n})$ with bounded error, polynomial space and no quantum random access memory. An algorithm that decides 4-colourability is a correct $\COL(\cdot, 4, \mu)$ for every $\mu$, so the bound $f_4$ may replace $d_4(\mu)$ when it is smaller. The depth of the recursion is at most $k \le 20$, so the factors from nested search and error reduction stay polynomial.

\subsection{Evaluation as an upper bound}

The maximum over $\delta$ in~\eqref{eq:r2} has no closed form, and $d_{k'}$ and $d_{k-k'}$ are needed at many arguments. We replace both by quantities that the algorithm itself can use.

Fix a grid $\mu_j = j/M$ for $j = 0, \dots, M$, and let $\lceil\mu\rceil_M$ be the smallest grid point at least $\mu$, for $\mu \le 1$. For each range $[k'/k, \min(1, \mu_j k')]$, fix a partition into $P$ pieces $[a, b]$. Define $\hat d_1 = \hat d_2 = 0$, $\hat d_3 = f_3$, and for $k \ge 4$ let $\hat d_k(\mu_j)$ be the minimum of $f_4$ if $k = 4$, of line~\eqref{eq:r1} with $\hat d_{k-1}(1)$ in place of $d_{k-1}(1)$ and of
\begin{equation}\label{eq:hat}
\min_{2 \le k' \le k/2}\; \max_{[a,b]}\; \frac{\max_{\delta \in [a,b]} h(\delta)}{2} + \max\Bigl(b\, \hat d_{k'}\bigl(\lceil \min(1, \tfrac{\mu_j}{a}) \rceil_M\bigr),\; (1-a)\, \hat d_{k-k'}\bigl(\lceil \min(1, \tfrac{b}{k'(1-b)}) \rceil_M\bigr)\Bigr),
\end{equation}
where the argument of $\hat d_{k-k'}$ is $1$ when $b = 1$.

\begin{proposition}\label{prop:hat}
For every $k \ge 1$ and grid point $\mu_j \ge 1/k$, there is a quantum procedure $\COL(\cdot, k, \mu_j)$ with bounded error, polynomial space and no quantum random access memory, in time $O^*(2^{\hat d_k(\mu_j) n})$.
\end{proposition}

\begin{proof}
By induction on $k$. For $k \le 2$ colouring takes polynomial time, and for $k = 3$ we use Lemma~\ref{lem:three} and the walk of~\cite{Mon18} and ignore $\mu$. For $k = 4$ we may also use Theorem~\ref{thm:four} and ignore $\mu$. Let $k \ge 4$. Reduction~1 calls $\COL(\cdot, k-1, 1)$, and $1 = \mu_M$ is a grid point, so its time is line~\eqref{eq:r1} with $\hat d_{k-1}(1)$.

For Reduction~2 with a fixed $k'$, every size $t$ in the search has $\delta = t/n$ in some piece $[a,b]$. For $t$ in this piece we call
\[
\COL\bigl(G[S], k', \lceil \min(1, \mu_j/a) \rceil_M\bigr) \quad\text{and}\quad \COL\bigl(G - S, k - k', \lceil \min(1, b/(k'(1-b))) \rceil_M\bigr).
\]
These limits are at least the ones of Reduction~2: since $t \ge an$, the classes of $G[S]$ have size at most $\mu_j n \le (\mu_j/a)\, t$, and since $\delta \le b$, the remaining classes have size at most $t/k' \le (b/(k'(1-b)))(n-t)$. A limit of $1$ always holds, and it is the one used when $b = 1$, where $G - S$ can still have up to $(1-a)n$ vertices. So the procedure stays correct. These limits lie in the domain of $\hat d$: since $a \le \mu_j k'$ we have $\mu_j/a \ge 1/k'$, and since $b \ge k'/k$ we have $b/(k'(1-b)) \ge 1/(k-k')$. For fixed $k$, $M$ and $P$ there are finitely many pieces and grid points, so all polynomial factors are uniform. By induction the two calls take time $O^*(2^{\hat d_{k'}(\cdot)\, t})$ and $O^*(2^{\hat d_{k-k'}(\cdot)(n-t)})$ at the stated grid points, and $t \le bn$ and $n - t \le (1-a)n$. Quantum search over the $\binom{n}{t} = O^*(2^{h(\delta)n})$ sets $S$ of size $t$ costs a factor $2^{h(\delta)n/2} \le 2^{n \max_{[a,b]} h/2}$. Summing over the polynomially many $t$ and taking the best $k'$ gives~\eqref{eq:hat}. The procedure is run with $k$ and the reduction fixed in advance, and the base cases have a running time that depends only on $n$, so the circuit is the same for every input of size $n$.
\end{proof}

A call $\COL(G, k, 1)$ decides $k$-colourability, so $\hat d_k(1)$ is an exponent for $k$-colouring.

\subsection{The values}

Take $f_3 = \frac12 \log \beta$ from Lemma~\ref{lem:three}, and $f_4 = \frac12 \log 1.6981$ from Theorem~\ref{thm:four}, or $f_4 = \frac12 \log 1.7159$ for the column with~\cite{WGJ24} alone.

\paragraph{Rows $k = 4, 5$.} Theorem~\ref{thm:four} gives the row $k = 4$. For $k = 5$ line~\eqref{eq:r1} suffices. Its term for $s$ equals $\log$ of $s^{1/(2s)} x^{(s-1)/s}$ with $x = 2^{d_4(1)}$, and this is at most $\log y$ whenever $s\, x^{2(s-1)} \le y^{2s}$. With $x = \sqrt{1.6981}$ and $y = 1.4518$, and with $x = \sqrt{1.7159}$ and $y = 1.4578$, these are polynomial inequalities for $s = 1, \dots, 5$, all true.

\paragraph{Rows $k = 6$ to $20$.} We evaluate~\eqref{eq:hat} with $M = P = 4000$, equal pieces, and add $10^{-9}$ to each exponent, which is far above the floating point error. Table~\ref{tab:results} shows the results, rounded up. The optimal $k'$ at $\mu = 1$ is $\lfloor k/2 \rfloor$ for $k = 6, \dots, 12$, and $6, 6, 7, 7, 7, 8, 8, 8$ for $k = 13, \dots, 20$.

As a test of the evaluation, the same computation without $f_4$ gives the precise values of Tables~2 and~3 of~\cite{SM22} to within $0.0008$ in the base, and in all rows at least these values, as expected from an upper bound.

\section{Domination problems}\label{sec:dom}

A dominating set of a graph is a set $D$ of vertices such that every vertex is in $D$ or has a neighbour in $D$. An independent dominating set is a dominating set with no edge inside it, and an edge dominating set is a set of edges such that every edge shares an endpoint with one of them. Minimum Maximal Matching asks for a smallest maximal matching. In each case the decision version asks for a solution of size at most $k$.

\begin{table}[h]
\centering
\small
\renewcommand{\arraystretch}{1.2}
\begin{tabular}{@{}p{4.3cm}p{7.6cm}l@{}}
\toprule
Problem & Previous best bound & This paper\\
\midrule
Dominating Set & $2^{n/2}$ quantum~\cite{WCCV23}; $1.3100^n$ quantum from~\cite{FGPS08}; $1.4864^n$~\cite{Iwa11}, $1.4969^n$~\cite{RB11} classical & $\mathbf{1.2235^n}$\\
Independent Dominating Set & $1.3351^n$~\cite{BDEP13}, $2^{0.424n}$~\cite{BEP10} classical; $1.2010^n$ quantum & $\mathbf{1.1583^n}$\\
Edge Dominating Set & $1.3160^n$~\cite{XN14}, $1.3226^n$~\cite{RB12} classical; $1.2010^n$ quantum & $\mathbf{1.1501^n}$\\
Minimum Maximal Matching & $1.3160^n$ classical~\cite{XN14,RB12}; $1.2010^n$ quantum & $\mathbf{1.1501^n}$\\
Weighted versions of the last two & $1.3226^n$ classical~\cite{RB12}; $1.2010^n$ quantum & $\mathbf{1.1501^n}$\\
\bottomrule
\end{tabular}
\caption{Domination problems on $n$ vertices. The classical bounds of~\cite{Iwa11,RB11,RB12,BEP10} are for polynomial space, and the previous bounds are the best known to us. The quantum bounds are for bounded error, polynomial space and no quantum random access memory, up to polynomial factors. The quantum value $1.2010^n$ is the walk on the branching enumeration of maximal independent sets, with at most $3^{n/3}$ leaves~\cite{MM65}, and for edge domination of the minimal vertex covers; to our knowledge it has not been stated. The value $1.3100^n$ is the walk on the enumeration of minimal dominating sets, if that enumeration is a backtracking tree. The bound $1.3160^n$ for Minimum Maximal Matching is implied by~\cite{XN14} through the conversion in~\cite{RB12}. For unweighted graphs, Minimum Maximal Matching and Edge Dominating Set have the same optimum, with a conversion in polynomial time~\cite{RB12}.}
\label{tab:dom}
\end{table}

\begin{theorem}\label{thm:dom}
With bounded error, polynomial space and no quantum random access memory, Dominating Set has a quantum algorithm in time $O^*(1.4969^{n/2}) \subset O(1.2235^n)$, Independent Dominating Set one in time $O(1.1583^n)$, and Edge Dominating Set, Minimum Maximal Matching and their weighted versions with non-negative integer weights of polynomial bit length have quantum algorithms in time $O^*(1.3226^{n/2}) \subset O(1.1501^n)$.
\end{theorem}

\begin{proof}
For each problem we show that the decision version of the classical algorithm, with a budget $k$, is a backtracking tree. Each step of these algorithms has the form $\mathrm{opt}(x) = \min_i (c_i + \mathrm{opt}(x_i))$ with values $c_i$ computed in polynomial time, which is equivalent to the OR of the decisions on $x_i$ with budget $k - c_i$. So each node answers, makes one call with a new budget, or branches into the OR of polynomially many children; it computes them in polynomial time and stores an instance of polynomial size, and the depth is polynomial. A node is labelled by its path of child indices in binary, so that the walk sees a tree of constant degree, also where two paths reach the same instance. The classical analysis bounds the number of leaves of the full search tree by $p(n)\,c^n$ for the classical base $c$ and a fixed polynomial $p$, and the tree for a budget $k$ is a subtree of it. The depth is at most $2n$ for Dominating Set, since each step removes a set or an element and the rule below removes three sets and two elements and adds one set; at most $n$ for edge domination, since each step lowers the number of undecided vertices; and at most $2n + n^2$ for Independent Dominating Set, counting vertices, free vertices and edges. So the tree has at most $T_n = (2n + n^2 + 1)\,p(n)\,c^n$ nodes. The walk of~\cite{Mon18} with the bound $T_n$ then decides the problem in time $O^*(c^{n/2})$, the same for every input of size $n$. Trying every budget, or a binary search for integer weights, gives the optimum. A solution follows by descending the tree from the root, running the detection on each child and moving to one that contains a marked leaf, then replaying the path and undoing each reduction as in the sources, in polynomial time; with errors amplified to $1/\mathrm{poly}(n)$ per call this costs a polynomial factor.

\emph{Dominating Set.} Algorithm~4 of~\cite{RB11}, in the numbering of its technical report, works on the set cover instance with the closed neighbourhoods as sets over the vertices. If no element is left, it answers yes when $k \ge 0$; if $k \le 0$ or an element lies in no set, it answers no; if all sets have at most two elements, a maximum matching gives the minimum cover, which it compares with $k$. Its reduction rules make one call each. An element of frequency one forces its set, with budget $k-1$. A set contained in another set, and an element whose sets include all sets of another element, are removed, with budget $k$. The counting rule of~\cite{RB11} takes a set, with budget $k-1$. The rule for a set $R = \{e_1, e_2\}$ of two elements of frequency two, with second sets $R_1$ and $R_2$, replaces $R, R_1, R_2$ by $(R_1 \cup R_2) \setminus R$ on the elements outside $R$. Since the subset rules are applied first, $R_1 \ne R_2$ and $R \not\subseteq R_i$. By the argument in~\cite{RB11}, the minimum cover then drops by exactly one, so the call has budget $k - 1$. Otherwise the algorithm branches on a set $S$ of maximum size, into taking $S$ with budget $k-1$ or discarding $S$ with budget $k$. Its running time is $O^*(1.4969^n)$ in polynomial space~\cite{RB11}, and $\sqrt{1.4969} \le 1.2235$.

\emph{Independent Dominating Set.} The algorithm of~\cite{BEP10} keeps vertices free or marked; a marked vertex may not enter the solution and must still be dominated. For a chosen vertex $v$, some free $u \in N[v]$ is in the solution, and the algorithm branches over these $u$, removes $N[u]$ and lowers the budget by one, marking the vertices tried in earlier branches. Its other steps have the same form: binary branches that take a vertex or mark it, branches that take several vertices together, forced additions, removal of equivalent vertices and of edges between marked vertices, and prunings of branches that cannot improve the solution. The analysis uses a measure of at most $n$ that is a sum of non-negative vertex weights. Some branchings are analysed there together with the branching that follows on a marked vertex of degree at most two created by the first one. We let the algorithm branch next on that vertex and count the two steps as one node. With this, the cases of the analysis give factor at most $1.3413 < c_0 = 2^{0.424} < 1.34165$ at every node other than a branching on a vertex of minimum degree in a regular component with degree $\delta \in \{2, 3, 4, 5\}$, which we call bad. The recurrence printed in the statement of Lemma~6 of~\cite{BEP10} has factor $1.4051$, while each case in its proof has factor at most $1.3413$. A vertex of minimum degree with a neighbour of larger degree is preferred, so all minimum-degree vertices lie in regular components at a bad branching. Remark~1 of~\cite{BEP10} counts bad branchings as finitely many. On a disjoint union of regular components they can occur once per component, so we complete the argument with one more step. Fix $K = 10^7$. A connected component $H$ with at most $K$ vertices is solved directly in constant time, and the algorithm makes one call on the rest with budget $k - \mathrm{opt}(H)$, or answers no if $H$ has no solution. If this step removes the component of the marked vertex between the two steps of a combined node, that branch drops by the whole measure of the component, which is at least the drop in each child of the second step, since those children change only vertices of the component. So the combined node keeps its bound. Now take two bad branchings on a path from the root, with components $H$ and later $H'$. No vertices are added and degrees do not increase, and the first branching removes a vertex of $H$. So either $H'$ is disjoint from $H$, or $H'$ is a proper subset of the connected graph $H$, in which case it has smaller degree. These components therefore form a laminar family of depth at most $4$. Each of them has more than $K$ vertices, since smaller components are removed first, and the members at the same height in the family are disjoint. So a path has at most $4n/K$ bad branchings. A bad branching has at most $\delta + 1 \le 6$ children and the measure does not increase, so its sum $\sum_i c_0^{-\mathrm{drop}_i}$ is at most $6$, and every other node, combined as above, has sum at most $1$. Giving each leaf the product along its path of $c_0^{-\mathrm{drop}_i}$ divided by the sum at the node, the weights of the leaves add up to one, and each weight is at least $c_0^{-n}\, 6^{-4n/K}$. So there are at most $(c_0\, 6^{4/K})^n$ leaves, and $1.34165 \cdot 6^{1/2500000} \le 1.1583^2$. The algorithm with this extra step uses polynomial space.

\emph{Edge Dominating Set and Minimum Maximal Matching.} Algorithm~4 of~\cite{RB12}, in the numbering of its technical report, splits the vertices into a vertex cover $C$, an independent set $I$ and undecided vertices $U$. A clique component $H$ of $G[U]$ of any size is handled by one call on the graph with a new vertex $v$ joined to $H$, with $C \cup H \cup \{v\}$ and $U \setminus H$, whose optimum is exactly one larger, so the call has budget $k + 1$. The new vertices keep the graph at most $2n$ vertices. Otherwise it branches on an undecided vertex, into $I$ with its undecided neighbours into $C$, or into $C$, and on the paths and cycles of $G[U]$ into two or a constant number of children. At a leaf $C$ is a vertex cover, and the smallest edge dominating set whose endpoints include $C$ is computed through a maximum matching. Unweighted Minimum Maximal Matching has the same decision problem by the conversion in~\cite{RB12}. For the weighted versions, Algorithm~5 of~\cite{RB12} replaces the clique rule by its Rule~2, which puts isolated undecided vertices into $I$ with budget $k$, and its Rule~3, one call on cliques of two or three vertices with the budget raised by the weight of the new edge, and larger cliques are branched on. Algorithm~7 of~\cite{RB12} replaces the clique rule by its Rule~4, one call on a clique component of any size with zero-weight new edges and budget $k$. Their leaves are a generalized edge cover or a perfect matching, both computed in polynomial time. The running time is $O^*(1.3226^n)$ in polynomial space, and $\sqrt{1.3226} \le 1.1501$.
\end{proof}

The faster algorithms of~\cite{Iwa11} for Dominating Set, of~\cite{BDEP13} for Independent Dominating Set and of~\cite{XN14} for Edge Dominating Set and unweighted Minimum Maximal Matching would give the bounds $O(1.2192^n)$, $O(1.1555^n)$ and $O(1.1472^n)$ by the same argument if they are backtracking trees in this sense. Their structure is not covered here.

\section{Monotone local search}\label{sec:mls}

Monotone local search~\cite{FGLS19} turns an algorithm that extends a partial solution into an exact algorithm. An implicit set system $\Phi$ maps an instance $I$ of size $N \ge n$ to a universe $U_I$ of $n$ elements and a family $F_I$ of subsets, with membership decidable in polynomial time. The problem $\Phi$-Subset asks for a set in $F_I$. The problem $\Phi$-Extension asks, for $X \subseteq U_I$ and an integer $k$, whether some $S \subseteq U_I \setminus X$ with $|S| \le k$ has $X \cup S \in F_I$. If $\Phi$-Extension is solved in time $c^k N^{O(1)}$, then $\Phi$-Subset is solved by a randomized algorithm in time $(2 - 1/c)^n N^{O(1)}$~\cite{FGLS19}.

\begin{theorem}\label{thm:mls}
Let $c > 1$, and suppose that $\Phi$-Extension is decided by a backtracking tree with at most $p(N)\,c^k$ nodes, for a fixed polynomial $p$. Then $\Phi$-Subset has a quantum algorithm with bounded error, polynomial space and no quantum random access memory, in time $(2 - 1/c)^{n/2} N^{O(1)}$. The same holds in the permissive form where the tree, on $(I, X, k)$, accepts whenever some $S$ with $|S| \le k$ has $X \cup S \in F_I$, and each accepting leaf gives a set $S$ of any size with $X \cup S \in F_I$.
\end{theorem}

\begin{proof}
Fix $k' \le n$, and let $t \le k'$ minimize $\binom{n}{t}\binom{k'}{t}^{-1} c^{k'-t}$. Prepare the uniform superposition over the subsets $Y \subseteq U_I$ of size $t$, which takes polynomially many gates. On each $Y$, run the walk of~\cite{Mon18} on the tree of $(I, Y, k'-t)$ with the bound $T = p(N)\,c^{k'-t}$, with its error reduced to $\delta$ by a majority over $O(\log 1/\delta)$ runs and with measurements deferred: compute the answer, copy it into one qubit, and uncompute. This oracle takes time $\sqrt{T}\,\poly(N)$, the same for every $Y$. Let $p_0 = \binom{k'}{t}\binom{n}{t}^{-1}$ and $m = \lceil p_0^{-1/2} \rceil$. Measure once without amplification and once after $j$ rounds of amplitude amplification~\cite{BBHT98,BHMT02}, for $j$ uniform in $\{0, \dots, m-1\}$, and accept if a measured $Y$ passes the extension decision again, with a majority over $O(\log n)$ runs. Do this for every $k' \le n$. The walk needs an explicit bound $T$ on the nodes of the full tree, explored without stopping at an accepting leaf.

If $F_I$ contains a set $S$ of size $k'$, every $Y \subseteq S$ with $|Y| = t$ is a yes-instance of the extension, with $S \setminus Y$ as completion, so with an exact oracle the success probability $p = \sin^2\theta$ of the start state is at least $p_0$. If $p > 3/4$, the measurement without amplification succeeds with probability above $3/4$. Otherwise $\sin 2\theta = 2\sqrt{p(1-p)} \ge \sqrt{p} \ge \sqrt{p_0}$, so $m \sin 2\theta \ge 1$. Then
\[
\frac{1}{m}\sum_{j<m} \sin^2((2j+1)\theta) \;=\; \frac12 - \frac{\sin(4m\theta)}{4m\sin 2\theta} \;\ge\; \frac14 .
\]
An oracle with error $\delta$ differs from the exact one by $O(\sqrt{\delta})$ in norm per call, so $\delta = 1/(m^2\poly(n))$, at a cost of $O(\log m) = O(n)$ repetitions, changes this by at most $1/100$. If $F_I$ is empty, no sample is marked, also in the permissive form, and the final check rejects except with small probability. A constant number of repetitions gives bounded error.

The run for $k'$ takes time $m\sqrt{T}\,\poly(N) = (\binom{n}{t}\binom{k'}{t}^{-1} c^{k'-t})^{1/2}\poly(N)$, the square root of the classical run in~\cite{FGLS19}. By Lemma~2.2 of~\cite{FGLS19}, $\max_{k'}\min_t \binom{n}{t}\binom{k'}{t}^{-1}c^{k'-t} \le (2-1/c)^n n^{O(1)}$, so the total time over all $k' \le n$ is $(2-1/c)^{n/2}N^{O(1)}$. The space is that of the walk and of $n$ sample qubits. A solution follows from an accepted $Y$ by descending the extension tree with the walk at each child and reading it off the accepting leaf. In the exact form, if the leaves do not give $S$, use self-reduction: stop if the current set is in $F_I$, and otherwise for $v$ outside the current set $Y$, run the decision on $(I, Y \cup \{v\}, b - 1)$ with the remaining budget $b$ and keep the first $v$ accepted.
\end{proof}

\begin{table}[h]
\centering
\small
\renewcommand{\arraystretch}{1.2}
\begin{tabular}{@{}p{3.6cm}p{3.4cm}lll@{}}
\toprule
Problem & Extension algorithm & $c$ & Classical, same extension & This paper\\
\midrule
Feedback Vertex Set & branching~\cite{IK19} & $3.460$ & $1.7110^n$ & $\mathbf{1.3081^n}$\\
Cluster Vertex Deletion & branching~\cite{Tsu19} & $1.811$ & $1.4479^n$ & $\mathbf{1.2033^n}$\\
Subset and Group Feedback Vertex Set & LP-branching~\cite{IWY16} & $4$ & $1.7500^n$ & $\mathbf{1.3229^n}$\\
Node Unique Label Cover, fixed labels $\Sigma$ & LP-branching~\cite{IWY16} & $|\Sigma|^2$ & $(2-|\Sigma|^{-2})^n$ & $\mathbf{(2-|\Sigma|^{-2})^{n/2}}$\\
3-Hitting Set, also weighted & branching~\cite{SZ15} & $2.168$ & $1.5388^n$ & $\mathbf{1.2405^n}$\\
Proper Interval Vertex Deletion & branching~\cite{Cao15} & $6$ & $1.8334^n$ & $\mathbf{1.3541^n}$\\
Multicut on Trees & branch and search~\cite{KLL13} & $\sqrt{\sqrt2+1}$ & $1.3565^n$ & $\mathbf{1.1647^n}$\\
Min-Ones $d$-SAT, $d \ge 3$; $d$-Hitting Set, $d \ge 4$ & clause branching~\cite{FGLS19} & $d$ & $(2-1/d)^n$ & $\mathbf{(2-1/d)^{n/2}}$\\
\bottomrule
\end{tabular}
\caption{Corollaries of Theorem~\ref{thm:mls}. For Multicut on Trees, $n$ is the number of edges. For $d = 3, 4, 5$ the last row gives $1.2910^n$, $1.3229^n$ and $1.3417^n$. The classical column applies~\cite{FGLS19} to the same extension algorithm; for $d$-Hitting Set~\cite{FGLS19} gives faster classical bounds through other extension algorithms, such as $1.5182^n$ for 3-Hitting Set.}
\label{tab:mls}
\end{table}

\begin{theorem}\label{thm:mlscor}
With bounded error, polynomial space and no quantum random access memory, the problems of Table~\ref{tab:mls} have quantum algorithms with the running times listed there, up to polynomial factors.
\end{theorem}

\begin{proof}
In each case, removing $X$ (deleting the vertices, cutting the edges or setting the variables to one and simplifying; for hitting sets, removing the hyperedges hit by $X$ and lowering the weight bound by the weight of $X$) turns the extension instance $(I, X, k)$ into an instance of the same problem with budget $k$, and we check that the algorithm is a backtracking tree whose number of nodes is bounded by its stated running time. The measure and conquer analyses prove a bound $C c^k$ on the leaves of the full tree by induction on the measure, for a constant $C$ that is hard-wired into the bound, and the depth is a fixed polynomial, which gives the explicit bound $T$.

\emph{Feedback Vertex Set.} The algorithm of~\cite{IK19} works on $(G, F, k)$ with undeletable vertices $F$, empty at the start. Its reductions delete a vertex, bypass a vertex of degree two or lower a multiplicity, each with one call of budget $k$ or $k-1$. A reduction to matroid matching answers in polynomial time when the maximum degree is at most three, and a pruning rule answers no. Otherwise it branches into deleting a vertex $u$ of maximum degree, with budget $k-1$, or contracting $u$ with its neighbours in $F$ into $F$, with budget $k$. Every branch removes a free vertex and every reduction shrinks the multigraph, so the depth is polynomial. The measure of~\cite{IK19} is $k$ at the start and nonnegative after pruning, so the full tree has $O^*(3.460^k)$ leaves, and $\sqrt{2 - 1/3.460} \le 1.3081$.

\emph{Cluster Vertex Deletion.} The reductions of~\cite{Tsu19} answer or make one call. Its branching rules branch over the twins of a vertex $v$ and over the members of a dominating family of vertex covers of an auxiliary graph, computed by a recursion that branches on a constant number of sets in each step. This family can be large, so we expand the recursion inside the tree: each of its leaves $X$ is followed by the call on $G - X$ with budget $k - |X|$, and the same is done for the second family built by one of the rules. Section~6 of~\cite{Tsu19} counts the calls of both the main recursion and the inner recursion, including inner lists with arbitrary covers, and bounds them by $O^*(1.811^k)$, and $\sqrt{2-1/1.811} \le 1.2033$.

\emph{Group and Subset Feedback Vertex Set, Node Unique Label Cover.} The algorithms of~\cite{IWY16} are LP-branching. Each node solves a half-integral relaxation in polynomial time and fixes, by single calls, assignments that do not raise the relaxed optimum. Otherwise it branches into two children for Group Feedback Vertex Set (Lemma~8 of~\cite{IWY16}) or $|\Sigma|$ children for Unique Label Cover (Lemma~1 and Corollary~4 of~\cite{IWY16}), and each branch lowers the gap between the budget and the relaxed optimum by at least $1/2$. So there are at most $4^k$ or $|\Sigma|^{2k}$ leaves. Subset Feedback Vertex Set is equivalent to its edge version, with the edges at the terminals as the edge subset, on the same graph and with the same budget~\cite{CPPW11}, and the edge version reduces to Group Feedback Vertex Set on the same graph with the same budget by Lemma~5 of~\cite{CPP16}, with the group $\mathbb{Z}_2^{|S|}$ for the edge subset $S$. Since this group is large, the relaxation is solved in the oracle model of Section~5.2 of~\cite{IWY16}. For Group Feedback Vertex Set $\sqrt{2 - 1/4} \le 1.3229$; for Unique Label Cover $|\Sigma|$ is fixed.

\emph{3-Hitting Set and Weighted 3-Hitting Set.} The algorithm of Appendix~B.1 of the arXiv version of~\cite{SZ15}, for positive integer weights of polynomial bit length, works on 3-uniform hypergraphs with a weight bound $W$ and a size parameter $k$, and some of its rules create smaller hyperedges inside a step. Its reductions answer, solving hypergraphs of degree at most two in polynomial time, or make one call, one of them with modified weights, and each branching rule returns the first of two answers that is not NIL. A non-NIL answer is a hitting set of weight at most $W$, and one is returned whenever a hitting set of weight at most $W$ and size at most $k$ exists, which is the permissive form of Theorem~\ref{thm:mls}, and the answer gives the set. Its measure $k - \alpha(G) + 1$ with $\alpha \le 0.87$ and branching numbers at most $2.168$ give $O^*(2.168^k)$ leaves. The family $F_I$ is the set of hitting sets within the weight bound. Unit weights give 3-Hitting Set, and $\sqrt{2 - 1/2.168} \le 1.2405$.

\emph{Proper Interval Vertex Deletion.} If $G$ has an induced subgraph from a finite family of graphs with at most six vertices, found in linear time, the algorithm of~\cite{Cao15} branches into deleting each of its vertices with budget $k-1$; otherwise the problem is solved in linear time. So there are at most $6^k$ leaves, and $\sqrt{2 - 1/6} \le 1.3541$. Interval Vertex Deletion is not included, since the algorithm of~\cite{Cao16} solves components and modules to optimality before it continues.

\emph{Multicut on Trees.} Cutting $X$ removes its edges and the requests they separate and leaves a forest, which the algorithm of~\cite{KLL13} handles, since it cuts edges itself. Its reductions are single calls, and those that use minimum vertex covers do so on auxiliary graphs of maximum degree two, in polynomial time. The branching rules make two or a constant number of children, the depth is polynomial, and the number of leaves is at most $\rho^k$ with $\rho = \sqrt{\sqrt2+1}$. Then $\sqrt{2-1/\rho} \le 1.1647$.

\emph{Min-Ones $d$-SAT and $d$-Hitting Set.} The algorithm of~\cite{FGLS19} picks an unsatisfied clause and branches over its at most $d$ variables set to one, with budget $k-1$, so there are at most $d^{k+1}$ nodes.
\end{proof}

Quantum search over all subsets takes time $2^{n/2}$. For Feedback Vertex Set, the walk on the algorithm of~\cite{Raz06} gives $O(1.3748^n)$, as stated for Maximum Induced Forest in~\cite{LU-MIF}. The walk on a classical exact algorithm with base $b$ gives $\sqrt{b}$ when that algorithm is a backtracking tree. The exact bases listed in~\cite{FGLS19} give at best $\sqrt{1.7347} > 1.3170$ for Feedback Vertex Set, $\sqrt{1.6181} > 1.2720$ for Cluster Vertex Deletion, $\sqrt{1.8638} > 1.3652$ for Subset Feedback Vertex Set and $\sqrt{1.8704} > 1.3676$ for 4-Hitting Set, and~\cite{FGLS19} lists no exact algorithm faster than brute force for Group Feedback Vertex Set, Node Unique Label Cover, Multicut on Trees and Min-Ones $d$-SAT, and for Proper Interval Vertex Deletion only $(2-\varepsilon)^n$ with $\varepsilon < 10^{-20}$. To our knowledge, no quantum form of monotone local search has been stated before. The classical column of Table~\ref{tab:mls} uses the same extension algorithms; faster classical bounds exist for some rows, such as $1.6097^n$ for Min-Ones 3-SAT~\cite{FGLS19}. For 3-Hitting Set the walk on the exact algorithm of~\cite{Wah07} would give $\sqrt{1.6278} > 1.2758$ if it is a backtracking tree, and the $2.0755^k$ extension algorithm of~\cite{Wah07} would give $1.2322^n$ through Theorem~\ref{thm:mls} if it is one. The faster $2.0409^k$ algorithm of~\cite{Tsu25} computes the optimum of a component by recursive calls before a tail call and is not a plain OR tree, so it is not used. Problems that reduce to 3-Hitting Set on the same universe, such as Feedback Vertex Set in Tournaments, get the bound $1.2405^n$ as well, although their own extension algorithms may give better bounds if they are backtracking trees.

\subsection{Approximation}\label{sec:amls}

Let $F_I$ be nonempty and upward closed, let $\mathrm{OPT}$ be the smallest size of a set in $F_I$, and call a set in $F_I$ of size at most $\beta\,\mathrm{OPT}$ a $\beta$-approximation. Approximate monotone local search~\cite{EKMNS22,EKMNS24} turns extension algorithms with ratio $\alpha$ and running time $c^k$ into exponential-time $\beta$-approximations. Let $L$ be a finite list of pairs $(\alpha, c)$ with $\alpha, c \ge 1$ and let $\beta \ge 1$. For a guess $k$ of $\mathrm{OPT}$, a pair $(\alpha, c)$ and $t \le \beta k$, let $x = \max(0, \lceil (1 - \beta/\alpha)k + t/\alpha \rceil)$ and $\ell = k - x$, and let $h(n,k,t,x)$ be the probability that a uniform $t$-subset of $[n]$ meets $[k]$ in at least $x$ elements. Let $\mathrm{best}(L, \beta)$ be the base of $\max_{k \le n/\beta} \min_{(\alpha,c),t} c^\ell / h(n,k,t,x)$, up to polynomial factors. This is a variant of the bound in Equations~(5) and~(6) and Theorem~2.3 of~\cite{EKMNS24}, which agrees with it up to a constant factor in the cost.

\begin{theorem}\label{thm:amls}
Let $\beta \ge 1$, and let each pair $(\alpha, c) \in L$ come with an extension procedure of one of two kinds: (a) a randomized polynomial-time algorithm with branching probabilities at least $1/\poly(N)$ and polynomial depth that, on $(I, X, \ell)$ such that some $S$ with $|S| \le \ell$ has $X \cup S \in F_I$, returns $Y$ with $X \cup Y \in F_I$ and $|Y| \le \alpha\ell$ with probability at least $q(N, \ell) \ge c^{-\ell}/p(N)$, where $q$ is computable in polynomial time and the bound still holds after the probabilities are rounded to $O(\log N)$ bits, at the loss of a constant factor; or (b) for $\alpha = 1$, a backtracking tree with at most $p(N)c^\ell$ nodes that accepts $(I, X, \ell)$ exactly when such an $S$ exists, and whose accepting leaves give one. Then there is a quantum $\beta$-approximation for $\Phi$ with bounded error, polynomial space and no quantum random access memory, in time $\mathrm{best}(L,\beta)^{n/2} N^{O(1)}$.
\end{theorem}

\begin{proof}
For each $k \le n/\beta$, take the pair and $t$ that minimize $c^\ell/h(n,k,t,x)$. Prepare the uniform superposition over the $t$-subsets $X$, together with the random bits of the procedure for kind (a), with its branching probabilities rounded. Mark a branch if $Z = X \cup Y \in F_I$ and $|Z| \le \beta k$ for kind (a), or if the detection walk accepts $(I, X, \ell)$ for kind (b), with its error reduced as in the proof of Theorem~\ref{thm:mls}. If $k = \mathrm{OPT}$ and $|X \cap \mathrm{OPT}| \ge x$, the rest of the optimum completes $X$ within $\ell$, and $|Z| \le t + \alpha\ell \le \beta k$, since $\alpha x \ge (\alpha - \beta)k + t$, or $x = 0$ and $t + \alpha k \le \beta k$. So the marked probability is at least $q_0 = h(n,k,t,x)$ times $q(N,\ell)$ for kind (a). With $m = \lceil 1/\sqrt{q_0} \rceil$, one measurement without amplification and one after a uniform number of rounds below $m$ succeed with probability at least $1/4$, as in the proof of Theorem~\ref{thm:mls}. For kind (b), $Y$ is found by descending the tree. The run costs $\sqrt{c^\ell/h(n,k,t,x)}\, \mathrm{poly}(N)$, the square root of the classical cost, and the smallest checked set over all $k$ is returned, together with $U_I$ as a candidate, which lies in $F_I$ by monotonicity and is a $\beta$-approximation if $\mathrm{OPT} > n/\beta$.
\end{proof}

The randomized branching algorithms of~\cite{KS20} for Vertex Cover and 3-Hitting Set are polynomial-time procedures of kind (a) that succeed with probability at least $(c^*+\varepsilon)^{-k}$ for every $\varepsilon > 0$ and large $k$, where the numerically evaluated bases $c^*$ are rounded up to the tabulated $c$. If the hypothesis holds only in this form, the proof gives $(\mathrm{best}(L,\beta)+\varepsilon)^{n/2}$ for every $\varepsilon > 0$, and since $\mathrm{best}$ is continuous in the bases, the rounded-up entries below still hold. Polynomial-time approximations are of kind (a) with $c = 1$: a maximal matching for Vertex Cover, a maximal family of disjoint hyperedges for $d$-Hitting Set, and the algorithms of~\cite{BBF99,ENZ00,ADFH23,vHV13,GVY97} for Feedback Vertex Set, Subset Feedback Vertex Set, Cluster Vertex Deletion, Proper Interval Vertex Deletion and Multicut on Trees, with ratios $2, 8, 2, 6, 2$. The trees of Theorem~\ref{thm:mlscor} are of kind (b), since they count their budget and accept only within it; for 3-Hitting Set we use unit weights and $W = \ell$, so that an answer other than NIL has size at most $\ell$. Table~\ref{tab:amls} gives the resulting bounds.

\begin{table}[h]
\centering\small
\begin{tabular}{lccccc}
\toprule
Problem & $\beta = 1.1$ & $\beta = 1.2$ & $\beta = 1.5$ & $\beta = 1.8$ & exact, quantum\\
\midrule
Vertex Cover & $\mathbf{1.0555^n}$ & $\mathbf{1.0312^n}$ & $\mathbf{1.0050^n}$ & $\mathbf{1.0003^n}$ & $1.0953^n$\\
3-Hitting Set & $\mathbf{1.1607^n}$ & $\mathbf{1.1135^n}$ & $\mathbf{1.0458^n}$ & $\mathbf{1.0211^n}$ & $1.2405^n$\\
4-Hitting Set & $\mathbf{1.2288^n}$ & $\mathbf{1.1861^n}$ & $\mathbf{1.1233^n}$ & $\mathbf{1.0933^n}$ & $1.3229^n$\\
Feedback Vertex Set & $\mathbf{1.2161^n}$ & $\mathbf{1.1752^n}$ & $\mathbf{1.0987^n}$ & $\mathbf{1.0315^n}$ & $1.3081^n$\\
Subset Feedback Vertex Set & $\mathbf{1.2288^n}$ & $\mathbf{1.1861^n}$ & $\mathbf{1.1233^n}$ & $\mathbf{1.0933^n}$ & $1.3229^n$\\
Cluster Vertex Deletion & $\mathbf{1.1318^n}$ & $\mathbf{1.1047^n}$ & $\mathbf{1.0676^n}$ & $\mathbf{1.0315^n}$ & $1.2033^n$\\
Proper Interval Vertex Deletion & $\mathbf{1.2558^n}$ & $\mathbf{1.2097^n}$ & $\mathbf{1.1404^n}$ & $\mathbf{1.1066^n}$ & $1.3541^n$\\
Multicut on Trees & $\mathbf{1.1034^n}$ & $\mathbf{1.0817^n}$ & $\mathbf{1.0525^n}$ & $\mathbf{1.0315^n}$ & $1.1647^n$\\
\bottomrule
\end{tabular}
\caption{Quantum $\beta$-approximation by Theorem~\ref{thm:amls}, with bounded error, polynomial space and no quantum random access memory, rounded up. The last column is the quantum exact bound of Table~\ref{tab:mls}, and for Vertex Cover the bound through Maximum Independent Set of~\cite{SengeVC}. For Multicut on Trees, $n$ is the number of edges.}
\label{tab:amls}
\end{table}

Every quantum exact algorithm is also a $\beta$-approximation, and every entry is below the quantum exact bound. To our knowledge, no quantum algorithm designed for exponential-time approximation of these problems has been stated before. Amplitude amplification over random sets of size $\beta \cdot \mathrm{OPT}$, the quantum form of the brute-force approximation of~\cite{EKMNS22}, gives $1.3097^n$, $1.2580^n$, $1.1769^n$ and $1.1360^n$ for the four ratios. The classical bounds of~\cite{EKMNS22,EKMNS24} are the squares of the table entries or smaller, since they also use extension algorithms not covered by Theorem~\ref{thm:amls}, such as the randomized $2.69998^k$ algorithm for Feedback Vertex Set.

The weighted form of approximate monotone local search~\cite{EKMNS23} groups the elements into classes of weights within a factor $\gamma = 1 + \delta/2$, puts all classes far below the heaviest class of the optimum into the sample, and runs the unweighted scheme in each of the $O(\delta^{-1}\log(n/\delta))$ remaining classes with one extension procedure. The quantum form samples in every class inside one amplitude amplification, and it has these parts. The marked probability is at least the product of the hypergeometric probabilities of the classes, for the correct guess of the class sizes $k_i$ of the optimum. Classes with $k_i > n_i/\zeta$ go entirely into the sample, where $\zeta = \beta/(1+\delta)$. The marking threshold is $\tau = w(W_k) + \zeta \sum_i \gamma^{i+1} k_i$, with $W_k$ the part of the sample that is fixed by the guess, so that $\tau \le \beta\,\mathrm{OPT}_w$ at the correct guess. This gives the square root of the classical cost, with a subexponential factor for the guesses. The extension procedure may be a tree of kind (b$'$): a permissive tree with a weight threshold, which accepts on $(I, X, \tau - w(X), \ell)$ whenever a set of size at most $\ell$ within the threshold exists, and whose answers stay within the threshold. The weighted Vertex Cover algorithm of~\cite{SZ15} in polynomial space, with base $1.381$, returns the first answer that is not NIL, so it is a tree of this kind, and its answers may be larger than $\ell$. With the parts above, this gives a weighted $1.1$-approximation for Vertex Cover in $O(1.0788^n)$ and, through Theorem~\ref{thm:mls}, exact weighted Vertex Cover in $O(1.1296^n)$; weighted 3-Hitting Set gets $O(1.1607^n)$ at $\beta = 1.1$. The exact weighted bound $O^*(1.238^n)$ cited in~\cite{EKMNS23} would give a smaller value, $\sqrt{1.238} < 1.1127$, if that algorithm is a backtracking tree.

\section{Verification}\label{sec:lean}

The Lean library, built on Mathlib with only the standard axioms, proves the inequality $s\, x^{2(s-1)} \le y^{2s} \Rightarrow s^{1/(2s)} x'^{(s-1)/s} \le y$ for $0 \le x' \le x$, the bounds $\sqrt{1.6981} \le 1.3032$, $\sqrt{1.7159} \le 1.3100$ and $\sqrt{1.32703} \le 1.1520$, and the rows $k = 5$ above. A script checks the certificate of Theorem~\ref{thm:four} in decimal arithmetic. Not formalized are the rules of~\cite{WGJ24}, the terminal algorithms, the backtracking structure, the quantum walk, Theorem~9 of~\cite{SM22}, Proposition~\ref{prop:hat} and the rows $k = 6$ to $18$.

A second script runs the 4-colouring tree. On random states of graphs of maximum degree 4, one branching step at the weights of Theorem~\ref{thm:four} has $\sum_j r^{\mu(P_j) - \mu(P)} \le 1$ in the real measure, and $P$ is colourable if and only if a child is, on 27463 states that reach the cases of~\cite{WGJ24} for maximum degree at most 4. On 400 random terminal instances, (T1), (T2) and (T3) agree with brute force, with nine local lists per triangle in place of the blocks, and the blocks of Lemma~5.3 of~\cite{YS26} are checked with one and two triangles. On 300 random graphs with at most 16 vertices, the whole tree agrees with brute force and the terminal sizes add up to at most $r^{\mu(G,\emptyset,\emptyset)}$. The tree for 3-colouring of~\cite{BE05}, the elimination of two-valued variables and the factor $e^{o(n)}$ are not run.

For the domination problems, the Lean files check $\sqrt{1.4969} \le 1.2235$, $\sqrt{1.4969^n} \le 1.2235^n$, $\sqrt{1.3226} \le 1.1501$ and $1.34165 \cdot 6^{1/2500000} \le 1.1583^2$. Scripts run the decision procedures of Section~\ref{sec:dom}: Algorithm~4 of~\cite{RB11} with all its rules, the basic branching of~\cite{BEP10} with marking and the rule for small components, and Algorithm~3 of~\cite{RB12}, which has the same kinds of steps as its Algorithm~4. On 400, 400 and 300 random graphs with at most 15, 13 and 10 vertices, the smallest accepted budget equals the optimum computed by brute force in every case. The count of bad branchings is also checked on disjoint unions of regular graphs. The scripts run the unweighted procedures; the refined branchings of~\cite{BEP10} and of Algorithm~4 of~\cite{RB12}, the weighted rules and the bounds on the tree size are not run.

For Section~\ref{sec:mls}, the Lean files prove the identity $2\sin x \sum_{j<m}\cos((2j+1)x) = \sin(2mx)$, the averaging bound $\sum_{j<m}\sin^2((2j+1)\theta) \ge m/4$ for $m\sin 2\theta \ge 1$, and the bases of Table~\ref{tab:mls}. A script evaluates $\max_{k}\min_t \binom{n}{t}\binom{k}{t}^{-1}c^{k-t}$ against $(2-1/c)^n$, which it does not exceed, for $n \in \{50, 200, 1000, 3000\}$ and each $c$ of the table. A second script runs the sampling and amplification with an exact classical oracle on random instances of 3-Hitting Set, Feedback Vertex Set and Cluster Vertex Deletion with up to 10 elements: clause branching, the branching of~\cite{IK19} with its matroid matching step replaced by further branching, and branching on induced paths with three vertices. On 120 instances each, the smallest accepted budget equals the optimum, at least $\binom{k'}{t}$ samples are marked at $k' = \mathrm{OPT}$, the larger of $M/N$ and the averaged success probability is at least $0.416$, where $M$ of $N$ samples are marked, and state-vector runs agree with the closed form. The walk, the bounded-error oracle, the algorithms of~\cite{Tsu19,KLL13,IWY16,SZ15,Cao15} and the tree sizes are not run. For Section~\ref{sec:amls}, a script computes $\mathrm{best}(L,\beta)$ from the large-deviation exponent of the hypergeometric tail, reproduces the values $1.2072$, $1.2099$, $1.2068$ and $1.4156$ of~\cite{EKMNS24} and $1.114$ of~\cite{EKMNS22}, and checks that the exact cost at $n = 1000$ is within a factor $(n+1)^2$ of $\mathrm{best}(L,\beta)^n$ for every entry of Table~\ref{tab:amls}; the maximization over $k/n$ is numerical.

A classical version of the colouring recursion, with OR in place of quantum search, is compared with brute force on random graphs with 5 to 9 vertices and $3 \le k \le 8$. Its base case for $k \le 3$ respects the size limits strictly, at $k = 4$ it may ignore the limit, as the algorithm does, and the limits passed on are kept tight or enlarged at random. With either reduction chosen at random, and with Reduction~2 only, it agrees with brute force on 300 graphs each. As a control, lowering the limit for $G - S$ by one makes 80 of 300 answers wrong, so the test does depend on the limits.

\section{Discussion}

The gain for $k = 4$ comes from the branching algorithm of~\cite{WGJ24} and the terminal algorithms of~\cite{YS26} and carries through the reductions of~\cite{Bys04,SM22}. Any further classical improvement of 3-colouring or 4-colouring by a branching algorithm in polynomial space carries over in the same way, through the same computation. The exponents for $k \ge 6$ are upper bounds for the explicit procedure of Proposition~\ref{prop:hat}. With finer grids and pieces they approach the values of the recursion~\eqref{eq:r1},~\eqref{eq:r2}, which we expect but do not prove.

For the domination problems the quantum bounds are the square roots of the classical bounds used. They follow once the decision version of each classical algorithm is seen to be a backtracking tree, and any faster classical algorithm of this kind gives a faster quantum one in the same way.

For monotone local search the quantum bound is the square root of the classical one whenever the extension algorithm is a backtracking tree. Extension algorithms with steps that compute the optimum of a part before a tail call, such as the $2.0409^k$ algorithm for 3-Hitting Set~\cite{Tsu25}, and the multivariate and sample-dependent forms of monotone local search, are not covered here. The same holds for approximate monotone local search, where randomized extension procedures that are not polynomial-time procedures, and deterministic approximate branching algorithms, are not covered.

\section*{Use of AI}

The proofs, the code and the Lean files were developed with Claude Opus~5.5 (Anthropic), which was also used to search the literature, to check the proofs and to draft the text. The author is responsible for the content.

\end{document}